# A quantitative study on the growth variability of tumour cell clones *in vitro*

C. Tomelleri*, E. Milotti[†,‡], C. Dalla Pellegrina*, O. Perbellini[§], A. Del Fabbro[†,‡], M. T. Scupoli[¶] and R. Chignola*[,‡]

*Department of Science and Technology, University of Verona, Strada le Grazie 15 - CV1, I-37134 Verona, Italy*

[†]*Department of Physics, University of Trieste, Via Valerio 2, I-34127 Trieste, Italy*

[‡]*National Institutes for Nuclear Physics, Via Valerio 2, I-34127 Trieste, Italy*

[§]*Department of Clinical and Experimental Medicine, University of Verona, c/o Policlinico G.B.Rossi, I-37134 Verona, Italy*

[¶]*Interdepartmental Laboratory for Medical Research (LURM), University of Verona, c/o Policlinico G.B.Rossi, I-37134 Verona, Italy*

**Corresponding author:**   Dr. Roberto Chignola
Dipartimento Scientifico e Tecnologico
Università di Verona
Strada Le Grazie 15 - CV1
I-37134 Verona, Italy
Tel. +39 045 8027953
Fax +39 045 8027952
e.mail roberto.chignola@univr.it

**Abstract**

*Objectives*: In this study, we quantify the growth variability of tumour cell clones from a human leukemia cell line. *Materials and methods*: We have used microplate spectrophotometry to measure the growth kinetics of hundreds of individual cell clones from the Molt3 cell line. The growth rate of each clonal population has been estimated by fitting experimental data with the logistic equation. *Results*: The growth rates were observed to vary among different clones. Up to six clones with a growth rate above or below the mean growth rate of the parent population were further cloned and the growth rates of their offsprings were measured. The distribution of the growth rates of the subclones did not significantly differ from that of the parent population thus suggesting that growth variability has an epigenetic origin. To explain the observed distributions of clonal growth rates we have developed a probabilistic model assuming that the fluctuations in the number of mitochondria through successive cell cycles are the leading cause of growth variability. For fitting purposes, we have estimated experimentally by flow cytometry the maximum average number of mitochondria in Molt3 cells. The model fits nicely the observed distributions of growth rates, however, cells in which the mitochondria were rendered non functional ($\rho^0$ cells) showed only a 30% reduction in the clonal growth variability with respect to normal cells. *Conclusions*: A tumor cell population is a dynamic ensemble of clones with highly variable growth rate. At least part of this variability is due to fluctuations in the number of mitochondria.

INTRODUCTION

The last decades have been characterised by intense research on the molecular genetics of cancer cells. A huge amount of knowledge has been massed on various molecular details of tumour initiation, promotion and progression, starting from the central discovery of oncogenes and multistep carcinogenesis (Weinberg 1989). More recently, some researchers have also endeavoured to understand the interplay among tumour cells and the surrounding microenvironment, the latter including normal cells and a plethora of extrinsic signals that both tumour and non-tumour cells exchange (Hanahan & Weinberg 2000; Fidler 2002). These signals can modify the individual tumour cells' behaviour acting, for example, on the expression of genes, and there is also growing evidence that normal cells exert epigenetic control on neoplastic behaviour (Barcellos-Hoff 2001). It is now clear that tumour initiation, promotion and progression is a rather complex mixture of events taking place at both the individual cell and at the cell population levels (Hanahan & Weinberg 2000; Barcellos-Hoff 2001; Fidler 2002), whereby normal cells, tumour cells and their microenvironments represent a non reducible ecosystem (see e.g. Chignola *et al.* 2006, and references cited therein).

The heterogeneity of the growth kinetics of individual tumours is a known biological fact observed both *in vitro* and *in vivo* (Laird 1969; Norton 1985; Chignola *et al.* 1995; Wilson 2001). Furthermore, studies on the individual growth of three-dimensional tumour cell cultures (spheroids) under controlled conditions have revealed that tumour cells are *per se* an adaptive, self-organising complex biosystem (Deisboeck *et al.* 2001) and show unexpected emerging growth behaviours such as oscillations of the spheroid volume (Chignola *et al.* 1999). These observations have important implications if one takes into account the well known relationships existing between kinetic heterogeneity of tumour growth and the outcome of antineoplastic treatments (Laird 1969; Norton 1985; Chignola *et al.* 1995; Wilson 2001). There is however a scarcity of studies on the quantitative aspects of tumour growth variability (with some notable exceptions, see e.g. Fidler 2002; Rubin 2005; and references cited therein) and on explanations of its origin at the cellular or subcellular level.

Recently, the speed of convergence towards the asynchronous state of two human leukemia cell populations has been measured by means of an elegant flow cytometry method (Chiorino *et al.* 2001). The results imply that the duration of the cell cycle varies through generations in these tumour cell populations and indeed the observed variance around the mean cell cycle length was rather high (coefficient of variation CV≈18% and 25% for Molt4 and Igrov cells, respectively). We reasoned that if the variability of cell cycle duration were a "sufficiently" stable trait - i.e. if the duration does not change too much at each cell division - then the variability should propagate to daughter cells from individual clones. Thus, cell populations from individual clones (to which we refer as clonal populations) should show different growth rates. However, measuring these growth rates would require an experimental method that fulfils the following conditions:

1. it should allow to measure the growth of small cell populations;

2. it should only minimally perturb cell growth;

3. for statistical purposes, it should be applied to measure the growth of a high number of clonal populations.

Conventional methods for the determination of cell population size, such as direct cell counting, ATP quantification or radioactive labelling, cannot therefore be used.

Starting from these considerations we set out to develop a spectrophotometric approach to measure the growth of cell populations deriving from individual clones of the Molt3 human leukemia cell lines. The growth rates of hundreds of clonal populations were measured and analysed within the framework of a probabilistic model. The results show that populations from individual tumour cell clones have different growth rates and that this variability is likely to have epigenetic origin. Attempts towards the identification of the epigenetic factor(s) involved are also described.

Cell variants for growth rate are continuously generated in a tumour population, and this endogenous variability might confer a greater fitness to the whole population under the pressure of selective environments.

MATERIALS AND METHODS

**Cell culture and cloning conditions**

Cells from the human T lymphoblastoid line Molt3 (ATCC number CRL-1552) were cultured in RPMI 1640 medium (Biochrom AG) supplemented with 2 mM glutamine (Sigma), 20 mg/l gentamycin (Biochrom AG), and 10% heat-inactivated foetal bovine serum (FBS, Biochrom AG). Cells were grown at 37° C in a humidified 5% $CO_2$ atmosphere in T75 culture flasks (Greiner Bio-One) and periodically diluted into fresh medium to avoid starvation.

For cloning purposes, cells from exponentially growing cultures were seeded into the wells of round-bottomed 96-well culture plates at the concentration of 0.3 cells/well into 100 µl complete medium. For each cloning experiments cells were dispensed in 5-10 plates (480-960 wells). It is well known that at these limiting dilutions, the distribution of cells into the wells of culture plates follows the Poisson statistics (Lefkovits & Waldmann 1979) and hence it is expected that 74% of the wells will contain no cells, 22% exactly one cell, and the remaining 4% two cells or more. Limiting dilution assays showed that, for Molt3 cells, the plating efficiency, i.e. the ratio between the number of wells scored as positive for cell growth and the number of expected wells, ranges between 0.5 and 0.7 (not shown). Thus, the choice of seeding the cells at the concentration of 0.3 cells/well was dictated by the conflicting needs of obtaining a high number of clonal populations from a reasonable number of culture plates.

**Growth assays**

The culture plates used in cloning assays were periodically monitored for cell growth using a microplate spectrophotometer (Powerwave, Bio-Tek Instruments) and raw data analysed with the dedicated KC4™ software. For each well, the absorption spectra in the wavelength range between 380 and 750 nm were collected and compared to those measured for culture wells containing the medium only. The method will be fully described in the following section. The method was validated in parallel assays where cell growth was also measured by direct cell counting and ATP determination. For both assays, cells were initially seeded in 96-wells culture plates at a concentration of 10.000 cells/ml in 100 µl complete medium. Cells were counted using a Burker chamber under a light phase-contrast microscope (Olympus IX51) upon dilution 1:2 with the vital dye trypan-blue to exclude death cells. ATP was determined by the luciferine/luciferase method using the CellTiter-Glo® Luminescent Cell Viability Assay kit (Promega) by following the manufacturer's procedures. The emitted light was measured using a microplate luminometer (FL$_X$ 800, Bio-Tek Instruments) and the data expressed in luminescence arbitrary units. For both assays, at each time-point the growth of 6 independent cultures was measured and averaged data compared to those evaluated by spectrophotometry for the same number of cultures grown under identical experimental conditions.

**Mathematical analysis of growth curves**

The growth data measured for each clonal population were fitted with the logistic equation:

$$y(t) = y_0 + \frac{a}{1 + \exp\left(-\frac{t - t_0}{b}\right)} \qquad (1)$$

where $y(t)$ is the population size measured at time $t$ and $y_0$, $a$, $b$ and $t_0$ are parameters that assume positive values. It should be noted that equation (1) describes a growth process characterised by an initial exponential phase with growth rate $r$ given by:

$$r = \frac{1}{b} \qquad (2)$$

with a saturation level $y_m$ given by:

$$y_m = \lim_{t \to \infty} y(t) = y_0 + a \qquad (3)$$

Equation (1) was automatically fitted to growth data from different clonal populations by means of a χ-squared minimisation algorithm written *ad hoc* in the Mathematica v.5.0 environment (Wolfram Research). We used standard statistical quantities, such as the standard deviation of estimated parameter values, to accept or reject the results.

**Isolation of mitochondria**

Mitochondria were isolated from $10^7$ cells samples. Cells collected by centrifugation were washed several times firstly with PBS and then with NKM buffer (0.13 M NaCl, 5 mM KCl, 7.5 mM $MgCl_2$, 1 mM Tris-HCl pH 7.4) at 4° C. Cell pellets were finally resuspended in 700 μl lysis buffer (10 mM KCl, 1.5 mM MgCl2, 1 mM PMSF, 1 mM DTT, 10 mM Tris-HCl pH 7.4) and kept on ice for 3 min. Samples were carefully homogenised using a glass Potter and 50 μl of a 2 M saccarose solution were added. Homogenates were centrifuged at 1200 g for 5 min and the pellets containing intact cells and nuclei were discarded. This procedure was repeated for three times. The supernatants were then centrifuged at 7000 g for 10 min to separate the protein and the microsomal fractions from that of mitochondria. The pellets formed by purified mitochondria were finally resuspended in CAB buffer (20 mM Na-HEPES, 0.12 mannitol, 80 mM KCl, 1 mM EDTA, pH 7.4) containing a mixture of protases inhibitors (Complete tablets, Roche) and stored on ice.

**Flow cytometry assays**

Cells and purified mitochondria were stained with 200 nM MitoFluor Red 589 fluorescent dye (Invitrogen) which is a specific cell membrane-permeable marker for mitochondria, and commonly used to measure the mitochondrial mass in intact cells. Both cells and mitochondria were incubated with the dye for 30 min at 37° C. Cells were then washed with PBS, whereas mitochondria with CAB.

Cells were also stained for 30 min at 37° C with 5 μM Vybrant® DyeCycle$^{TM}$ Green Stain (Invitrogen), a fluorescent dye developed to label the cell DNA in intact living cells.

Fluorescence was measured with a FACScalibur flow cytometer (Becton-Dickinson) equipped with both Ar and red diode lasers, and data were analysed using the CellQuest (Becton-Dickinson) software.

**Quantification of mitochondria**

The average number of mitochondria per cell <M> was estimated by flow cytometry. Cell populations and isolated mitochondria were labelled with the mitochondria-specific fluorescent probe (see above), and the mean fluorescence intensities of labelled and unlabelled samples - measured at the same sensitivity of the photomultiplier - were used to calculate <M> as follows:

$$<M> = \frac{MFI_C - MFI_{UC}}{MFI_M - MFI_{UM}} \qquad (4)$$

where *MFI* is the measured mean fluorescence intensity of the samples and the subscripts *C*, *UC*, *M* and *UM* refer to intact cells, unlabelled cells, isolated mitochondria and unlabelled isolated mitochondria samples, respectively.

The standard deviation of the average <*M*> was calculated from the coefficient of variation (CV) of measured fluorescence intensity distributions:

$$\sigma_{C,UC,M,UM} = \frac{MFI_{C,UC,M,UM} \cdot CV}{100 \cdot \sqrt{n}} \quad (5)$$

where *n* is the number of measured events. The standard deviation of <*M*> was then calculated by applying the usual equations from error propagation theory, which, in this case, reduce to the following equation:

$$\sigma_{<M>} = <M> \cdot \sqrt{\frac{\sigma^2_C + \sigma^2_{UC}}{(MFI_C - MFI_{UC})^2} + \frac{\sigma^2_M + \sigma^2_{UM}}{(MFI_M - MFI_{UM})^2}} \quad (6)$$

**Preparing $\rho^0$ cells from Molt3 cultures**

$\rho^0$ cells were obtained by culturing Molt3 cells in the presence of 50 ng/ml ethidium bromide (Armand *et al.* 2004). Since the treatment with ethidium bromide inhibits the replication of mitochondrial DNA, the mitochondrial DNA is progressively depleted and ATP production through oxydative phosphorylation blocked (Armand *et al.* 2004). Cells with non-functional mitochondria cannot survive in standard complete RPMI-medium, and culture media were therefore supplemented with 100 µg/ml pyruvate and 50 µg/ml uridine as described in (Armand *et al.* 2004). Culture media were refreshed at 3 days intervals. After 24 days, the cells were cloned as described above in the absence of ethidium bromide, and clonal populations expanded for further 18 days. The cells from each clone were then split in two populations and grown in parallel for 11 days in RPMI medium and medium containing pyruvate and uridine. Cells growing in pyruvate and uridine containing medium only were scored as true $\rho^0$.

$\rho^0$ cells were also subjected to PCR analysis throughout the culture procedure and the results were compared to those obtained with normal Molt3 cells. To this purpose the DNA was extracted using the Wizard Genomic DNA purification kit (Promega) and 50 ng of total DNA were subjected to PCR using mitochondrial specific primers (Holmuhamedov *et al.* 2002) and primers for a fragment of the nuclear β-actin gene as the control. The primers were as follows:

1. mitochondrial DNA specific primers (giving an amplicon of 312 bp, Holmuhamedov *et al.* 2002):

forward: 5'-ACAATAGCTAAGACCCAAACTGGG-3';

reverse[1]: 5'-CCATTTCTTGCCACCTCATGGGC-3'.

2. nuclear DNA specific primers (giving an amplicon of 311 bp within the β-actin gene):

forward: 5'-CATTGCTCCTCCTGAGCGCAAG-3';

reverse: 5'-GCTGTCACCTTCACCGTTCCAG-3'.

The amplification products obtained after 30 PCR cycles were analysed by electrophoresis onto 1% agarose gels.

**Descriptive statistics**

---

[1] The primers were designed based on the data published in (Holmuhamedov *et al.* 2002). The sequence of our reverse primer, however, corresponds to the complementary sequence given in (Holmuhamedov *et al.* 2002) and indeed sequence alignment analysis showed that the two primers described by those Authors incorrectly bind the same DNA strand. In addition, our sequence was reduced by 1 base to limit the difference in the melting temperatures between the forward and the reverse mitochondrial DNA specific primers.

Where indicated in the next section, we compared experimental data with the two-tailed Student *t* test. Probability threshold values *P*<0.05 or *P*<0.01 were considered to reject the null-hypothesis and hence to define the differences between the data as statistically significant or highly significant, respectively.

RESULTS

**Measuring cell growth by microplate spectrophotometry**

Fig.1A shows the mean spectra measured for 60 wells of a microplate containing an initial cell number of 1.000 cells/well in 100 µl of complete growth medium and, for comparison, the mean spectra measured for 60 wells containing medium only. Two peaks at 410 nm and 560 nm are visible and these correspond to the absorbance properties of the pH-sensitive dye phenol-red used in RPMI growth medium. The absorbance intensities of the two peaks varied in time as a consequence of cell growth and hence of the acidification of the medium due to lactic acid secretion by cells. Only a slight variation in the absorption spectra was measured for the wells containing growth medium only. At the 730 nm wavelength, the spectra were not influenced by the absorption of phenol-red. At 730 nm the optical density of the cell samples increased with time, and the measures were not dependent upon temperature at least within the range 25-40 °C (not shown).

To test whether this increase correlated with cell growth the average optical density at 730 nm measured for 6 cell populations was compared to two independent measures of cell growth, namely direct cell counting (Fig.1B) and ATP determination (Fig.1C). These were obtained in parallel experiments carried out under identical growth conditions.

The major difference between the three methods is that microplate spectrophotometry is not sensitive to cell death due to starvation taking place at the end of the growth assay. Dead cells are not immediately removed from cell cultures and hence they interfere with the optical measurements. However, as far as the growth rate is concerned, microplate spectrophotometry is well suited to carry out measurements (see Fig.1B), and indeed it has the following great advantages over the other methods:

1. since cell sampling is not necessary to carry out measurements, the growth of individual cell populations can be measured in time;

2. as a consequence of the above point, one can avoid to make replicates of the same cell culture;

3. measurements only minimally perturb cell cultures;

4. it takes just a few minutes to measure the growth occurring in 96 wells.

Thus the growth of individual populations from cell clones can be measured (Fig.1D) and data fitted with growth equation (1).

**Growth rate variability of individual cell clones**

Molt3 cells were cloned, and the growth of 201 clonal populations - from three independent experiments - was measured by microplate spectrophotometry. The growth rate of each clonal population was estimated by fitting data with equation (1). The growth rates of Molt3 clones distributed as shown in Fig.2 and the mean value was found to be $1.128 \pm 0.036$ days$^{-1}$.

The variability of growth rates among different clonal populations can have either a genetic or an epigenetic origin or both. We reasoned that if the origin were genetic then the trait should have been conserved in the offsprings of a given clone. Six clones from the cloning experiments of Molt3 cells were therefore selected at random and their individual growth rate is reported in Fig.2. Two of the six selected clones (named as follows: number of microplate followed by the well coordinate) had a growth rate below the mean value of the Molt3 population whereas the other 4 had a value above that. The six clones were expanded up to populations sizes of ≈$10^6$ cells and further cloned. The growth rates of the offsprings of the two clones with an initial growth rate below the mean value measured for Molt3 cells and those from the 4 clones with a growth rate value above were grouped and their distributions analysed. Grouping was considered to minimize the source of errors given by the small number of isolated and measured subclones. As it can be appreciated in Fig.2, these distributions were qualitatively similar to that shown by the parent population. Indeed, the mean growth rate calculated for 82 clonal populations from the clones with initial growth rate below the mean value of Molt3 cells, and that calculated for 161 clonal populations from the clones with initial growth rate above the mean value of Molt3 cells were not significantly different to that of the parent Molt3 population ($P>>0.05$ in both cases, Student's *t* test).

Thus, growth variability of individual tumour cell clones is not a genetically inherited trait; rather, it is likely to have an epigenetic origin.

**A theoretical hypothesis for the observed growth variability of clonal populations: mitochondria as the possible epigenetic cause**

We attempted to explain the observed distributions of growth rates (see e.g. Fig.2) from a probabilistic perspective. To do this, we supposed that some epigenetic factor (Ef) existed in the cells and had the following properties:

1. the growth rate of a clonal population is proportional to the average amount of the Ef;

2. the Ef is randomly distributed at mitosis to newborn cells;

3. the Ef must increase its size during the cells cycle up to a maximum value that approximately doubles its initial size at mitosis.

Condition 2. is required to allow the offsprings of a cell clone to randomly change their growth rates. The last condition has instead a mechanistic justification: if the Ef did not grow during the cell cycle, or conversely if it did grow too much, then the daughter cells would receive a progressively lower or higher amount of the Ef, respectively.

Mitochondria are known to provide the vast majority of energy to tumour cells grown *in vitro* under the form of ATP, and this energy is used by cells to accomplish the various tasks of molecular synthesis required to progress along the cell cycle. In addition, mitochondria - as well as other cell organelles - are known to be randomly partitioned to the two daughter cells at mitosis (Birky 1983) and are known to double their number during the cell cycle (Posakony *et al.* 1977; James & Bohman 1981). The latter biological property was further assayed experimentally here by flow cytometry (Fig.3A).

Thus mitochondria are good candidates as the Ef causing the growth variability of tumour cell clones, and the probabilistic model that we have developed on the basis of the above considerations writes as follows (see the Appendix for details):

$$P_{r(M)}(r) = \int_{M_{\min}}^{M_{\max}} \frac{k}{r^2 \sqrt{\frac{\pi M}{2}}} e^{-2\frac{\left(\frac{M}{2}-\frac{k}{r}\right)^2}{M}} \cdot \frac{1}{\sqrt{2\pi\sigma^2_M}} e^{-\frac{(M-M_0)^2}{2\sigma^2_M}} dM \qquad (7)$$

where $P_{r(M)}(r)$ is the probability density of growth rates $r$ distributions, $k$ is the (linear, see the Appendix) growth rate of mitochondria, $M_0$ is the maximum average number of mitochondria in the cells (i.e. the maximum average number of mitochondria before mitosis and ranging, among different cells, between $M_{\min}$ and $M_{\max}$) and $\sigma_M$ is the standard deviation that parametrize the fluctuations of $M$ around $M_0$. It should be noted that three parameters are considered in equation (7), namely k, $M_0$ and $\sigma_M$, two of which are correlated ($M_0$ and $\sigma_M$). To estimate the values of these parameters by fitting, an independent measure of $M_0$ or of $\sigma_M$ is therefore required.

For this reason, we estimated $M_0$ experimentally by flow cytometry (Fig.3B). It should be noted that for fitting purposes the knowledge of the exact maximum number of mitochondria in the cells is not required and that a rough estimate is sufficient. Cells and isolated mitochondria were labelled with a mitochondria-specific fluorescent probe and measurements were processed as described in the Material and Methods section (Fig.3B). These calculations allowed us to estimate the mean number of mitochondria in the Molt3 cell population that turned out to be $<M>$ = 225.7 ± 3.9, a number in agreement with values reported also by others (Robin & Wong 1988). This estimate, however, corresponds to the average number of mitochondria in the desynchronized cell population. $M_0$ was therefore computed by assuming that the growth of mitochondria is linear in time (see also Posakony *et al.* 1977; James & Bohman 1981), that cells are evenly distributed in their cycles and that, as confirmed experimentally (Fig.3A), mitochondria double their initial number before mitosis. Thus:

$$M_0 \approx \frac{4}{3} <M> = 300.9 \qquad (8)$$

The above value was substituted in equation (7) and this equation, now with only two free and independent parameters, was numerically solved and fitted to growth rate distributions from independent experiments with Molt3 cells (Fig.4). The fits showed good agreement between theory and experiments.

**Growth rate variability of $\rho^0$ cells**

The results shown in Fig.4 suggest that the observed growth variability of Molt3 clones is correlated to the variability in the number of mitochondria they possess and transmit to their offsprings. If this were the biological basis of the growth variability of individual clones, then Molt3 cells in which mitochondria are rendered non functional - the so-called $\rho^0$ cells - should grow with less variable kinetics with respect to normal Molt3 cells.

We obtained $\rho^0$ cells by culturing Molt3 cells in the presence of ethidium bromide as described in the Materials and Methods section and they were identified and selected on the basis of:

1. loss of mitochondrial DNA (Fig.5A);

2. dependence on uridine and pyruvate for survival (see Materials and Methods).

Molt3 $\rho^0$ cells were then expanded and cloned, and the growth kinetics of individual clonal populations were analysed as described for normal Molt3 cells. The distribution of growth rates measured for 41 clonal populations is shown in Fig.5B. With respect to that of normal Molt3 cells (see e.g. Fig.2), this distribution appears shifted to the left and this is due to the growth kinetics of $\rho^0$ cells which resulted approximately 2.1 times slower than that of normal Molt3 cells (mean ± SE growth rate of 1.13 ± 0.036 and of 0.53 ± 0.026 days$^{-1}$ for normal and $\rho^0$ cells, respectively). The relative variability of growth rates was approximately 30% lower for $\rho^0$ cells with respect to normal cells (CV 45.1 % and 31.5 % for normal and $\rho^0$ cells, respectively). Thus, we conclude that the fluctuations in the number of mitochondria:

1. in general, are correlated to the growth variability of individual clones;

2. in particular, determine approximately 30% of the growth variability.

DISCUSSION

The fact that a tumour population is a heterogeneous ensemble of cell variants for a given trait has long been recognised, starting from the pioneering works on the spontaneous transformation of embryo fibroblasts *in vitro* and on the individual metastatic potential shown by tumour cell clones *in vivo* (see e.g. Fidler 2002; Rubin 2005; and references cited therein). These traits have been observed to vary among successive cell generations, and the pressure of environmental selection on genetically-determined cell variants has been shown to play a pivotal role in allowing the onset of dominant - often more aggressive - variants (Fidler 2002; Rubin 2005). Thus, the growth of a tumour can be viewed as an evolutionary process whereby tumour cells, normal cells and environmental niches constitute a dynamical ecosystem (Hanahan & Weinberg 2000; Barcellos-Hoff 2001; Vineis 2003; Fidler 2002; Rubin 2005). As expected, both genetic and epigenetic forces act to sustain competition between tumour and normal cells (Hanahan & Weinberg 2000; Barcellos-Hoff 2001; Vineis 2003; Fidler 2002; Rubin 2005).

The duration of the cell cycle is recognised among the phenotypic traits that can vary in a tumour cell population (see e.g. Chiorino *et al.* 2001 and references cited therein), and indeed the heterogeneity of tumour growth kinetics is known to have serious implications for tumour treatment (Laird 1969; Norton 1985; Chignola *et al.* 1995; Wilson 2001). For this reason, the variability in cell cycle duration has been extensively investigated for tumour cells both experimentally and theoretically (see e.g. Chiorino *et al.* 2001 and references cited therein). In particular, statistical approaches have been used to model experimental data taken at the level of whole tumour cell populations, where inter-cell variability was indirectly estimated by the mathematical analysis of data obtained in pulse-chase experiments with radioactive-labelled DNA bases (see e.g. Wilson 2001) or, more recently, with fluorescent probes (Chiorino *et al.* 2001). The direct measurements of cell cycle duration at single cell level has obvious technical limitations. Thus, we attempted to approach the problem from an experimental perspective that lies somewhat in between the two extremes represented by whole-population and single-cell measurements. We have shown that:

1. the heterogeneity in the growth kinetics of clonal populations, i.e. of small populations from a single tumour cell clone, can be measured *in vitro*;

2. the growth variability among clonal populations is likely to be determined at the epigenetic level.

Although we could not determine how the stability of this trait through generations, the data show that clonal populations after 20 days of culture - i.e. approximately 20 generations - possess a high degree of variability in their growth kinetics. The unexpected result, therefore, is that the variability in cell cycle duration among individual clones appears to propagate to their offsprings. This has allowed us to measure different growth kinetics for different clonal populations, after all. Propagation of the trait, on the other hand, is unlikely to be caused by genetic forces since the growth variability of populations generated by subclones have the same variability as the parent population and hence the trait is not fixed in the progeny. This means that the fluctuations of the "epigenetic factor" must be represented by a stochastic process with memory.

The variability in the cell cycle kinetics has been assumed to stem from unequal cell size division or from fluctuations in the concentration of some cellular components (see e.g. Chiorino *et al.* 2001 and references cited therein), both processes taking place at mitosis. These considerations have led us to hypothesise a role for mitochondria, but the data obtained with $\rho^0$ cells show that the fluctuations in the number of mitochondria take account of 30% only of the growth variability of clonal populations. Thus, the most important epigenetic factor(s) behind growth variability of tumour cell clones still remains to be identified.

Our data, however, indicate that a tumour population is a dynamic ensemble of noisy cell units. It is worth noting that noise, under the form of fluctuations in the concentrations of key molecules, can drive the differential expression of genes in otherwise identical cells (McAdams & Arkin 1999). This form of noise might also lead to different cycle times for individual clones thus determining the variability of growth rates and perhaps of the growth saturation levels. In fact, an open question that has not been addressed here is that clonal populations also reached different saturation levels (see e.g. Fig.1D) and we did not observe correlation with the growth rates (not shown). A new testable hypothesis to explain the observed variability of both growth rate and saturation level of clonal populations is therefore required.

ACKNOWLEDGEMENTS

We wish to thank Prof. Massimo Delledonne and Dr. Alberto Ferrarini, both at the Department of Science and Technology, University of Verona, for their help and assistance during the execution of PCR assays.

The authors also thank Giancarlo Andrighetto, former professor of immunology but still master of rock climbing, for his enthusiastic support. In particular, R.Chignola wishes to dedicate to him the efforts that he put in the present paper in gratitude for the intense discussions that he has had over the years and that he is still having on the subjects of the work, discussions that frequently take place in rather unusual laboratories such as the Dolomites' walls or the underground labyrinths of deep caves in the Apuane Alps.

APPENDIX

**A probabilistic model of the growth rate variability of tumor cell clones**

Since the variability of growth rates is likely to have an epigenetic origin (see the Results section) we assumed that an epigenetic factor (Ef) is randomly distributed at mitosis, and that the amount of inherited Ef determines the cell cycle duration in the daughter cells. Given their biological significance, we identified mitochondria as a good candidate Ef, and hence the model refers to the number of mitochondria as the stochastic variable. However, it should be noted that the model might apply to other cellular organelles and/or molecular mechanisms provided the following assumptions holds:

1. during the cell cycle the Ef approximately doubles its size and the growth is linear with time. This assumption holds for mitochondria (Posakony *et al.* 1977; James & Bohman 1981);

2. the Ef can be represented by a stochastic variable and daughter cells inherit random amounts of the Ef. Mitochondria fulfil this assumption as they have been observed to be randomly partitioned at mitosis (Birky 1983);

3. at mitosis the Ef is partitioned at random between the two daughter cells with equal probability. Mitochondria fulfil also this assumption as they are known to be distributed to daughter cells with equal probability and their distribution appears to follow a binomial distribution;

4. cells are supposed to divide when Ef reaches a maximum value Ef (max), and for this reason the amount of Ef in each generation is independent from that in previous generations (and cell division erases all existing memory).

From assumption 3, if the mother cell has $M$ mitochondria at division, the probability of finding $m$ mitochondria in a given daughter cell is:

$$P_m = \binom{M}{m} \frac{1}{2^M} = \frac{M!}{m!(M-m)!} \frac{1}{2^M} \tag{A1}$$

It is well-known that when $M \gg 1$, the binomial distribuion (A1) is well approximated by a Gaussian distribution with mean $M$ and standard deviation $\sigma \approx \sqrt{M}/2$:

$$P_m \approx \frac{1}{\sqrt{2\pi\sigma^2}} e^{-\frac{\left(m-\frac{M}{2}\right)^2}{2\sigma^2}} = \frac{1}{\sqrt{\frac{\pi M}{2}}} e^{-2\frac{\left(m-\frac{M}{2}\right)^2}{M}} \tag{A2}$$

From assumption 1, if $m$ is the number of mitochondria at the start of the cell cycle, $M$ is the number of mitochondria at division and $k$ is the mitochondrial proliferation rate, then:

$$M = m + kT \tag{A3}$$

where $T=1/r$ is the cell doubling time and $r$ the corresponding growth rate. Now, using the standard formulas for the substitution of random variables in probability densities

$$P_m dm = P_m \left|\frac{dm}{dT}\right| dT = P_T(T) dT$$
$$P_T(T) dT = P_T\left(\frac{1}{r}\right) \left|\frac{dr}{dT}\right| dr = P_r(r) dr \tag{A4}$$

we find the probability density for the growth rate $r$:

$$P_r(r) \approx \frac{k}{r^2 \sqrt{\frac{\pi M}{2}}} e^{-2\frac{\left(\frac{M}{2}-\frac{k}{r}\right)^2}{M}} \tag{A5}$$

In equations (A1) to (A5) we have implicitly assumed a fixed $M$, but obviously the maximum number of mitochondria is itself subject to fluctuations about a mean value $M_0$: if we assume that the fluctuations of *M* are approximately Gaussian with variance $\sigma^2_M$, then the probability density (A5) is transformed into a new density

$$P_{r,M}(r,M) = \frac{k}{r^2\sqrt{\frac{\pi M}{2}}} e^{-2\frac{\left(\frac{M}{2}-\frac{k}{r}\right)^2}{M}} \cdot \frac{1}{\sqrt{2\pi\sigma_M^2}} e^{-\frac{(M-M_0)^2}{2\sigma_M^2}} \tag{A6}$$

which depends on the two random variables *r* and *M*. Now, using the terminology of Bayesian statistics, *M* is a "nuisance variable", and we eliminate it as usual by marginalizing the density (A6), i.e., by integrating out the nuisance variable *M*:

$$P_{r(M)}(r) \approx \int_{M_{\min}}^{M_{\max}} \frac{k}{r^2\sqrt{\frac{\pi M}{2}}} e^{-2\frac{\left(\frac{M}{2}-\frac{k}{r}\right)^2}{M}} \cdot \frac{1}{\sqrt{2\pi\sigma^2_M}} e^{-\frac{(M-M_0)^2}{2\sigma^2_M}} dM \tag{A7}$$

where ($M_{\min}, M_{\max}$) is the range of *M*, and for all practical purposes we can take $M_{\max} - M_0 = M_0 - M_{\min} = n\sigma_M$, where $n \geq 3$.

REFERENCES

Armand R, Channon JY, Kintner J, White KA, Miselis KA, Perez RP, Lewis LD (2004) The effects of ethidium bromide induced loss of mitochondrial DNA on mitochondrial phenotype and ultrastructure in a human leukemia T-cell line (MOLT-4 cells). *Toxicol. Appl. Pharmacol.* **196**, 68-79.

Barcellos-Hoff MH (2001) It takes a tissue to make a tumor: epigenetics, cancer and the microenvironment. *J. Mammary Gland Biol. Neoplasia* **6**, 213-221.

Birky CW Jr (1983) The partitioning of cytoplasmic organelles at cell division. *Int. Rev. Cytol. Suppl.* **15**, 49-89.

Chignola R, Foroni R, Franceschi A, Pasti M, Candiani C, Anselmi, C, Fracasso G, Tridente G, Colombatti M (1995) Heterogeneous response of individual multicellular tumour spheroids to immunotoxins and ricin toxin. *Br. J. Cancer* **72**, 607-614.

Chignola R, Schenetti A, Chiesa E, Foroni R, Sartoris S, Brendolan A, Tridente G, Andrighetto G, Liberati D (1999) Oscillating growth pattern of multicellular tumour spheroids. *Cell Prolif.* **32**, 39-48.

Chignola R, Dai Pra P, Morato LM, Siri P (2006) Proliferation and death in a binary environment: a stochastic model of cellular ecosystems. *Bull. Math. Biol.* **68**, 1661-1680.

Chignola R, Milotti E (2005) A phenomenological approach to the simulation of metabolism and proliferation dynamics of large tumor cell populations. *Phys. Biol.* **2**, 8-22.

Chiorino G, Metz JAJ, Tomasoni D, Ubezio P (2001) Desynchronization rate in cell populations: mathematical modeling and experimental data. *J. Theor. Biol.* **208**, 185-199.

Deisboeck TS, Berens ME, Kansal AR, Torquato S, Stemmer-Rachamimov AO, Chiocca EA (2001) Pattern of self-organization in tumour systems: complex growth dynamics in a novel brain tumour spheroid model. *Cell Prolif.* **34**, 115-134.

Fidler IJ (2002) The organ microenvironment and cancer metastasis. *Differentiation* **70**, 498-505.

Hanahan D, Weinberg RA (2000) The hallmarks of cancer. *Cell* **100**, 57-50.

Holmuhamedov E, Lewis L, Bienengraeber M, Holmuhamedova M, Jahangir A, Terzic A (2002) Suppression of human tumor cell proliferation through mitochondrial targeting. *FASEB J.* **16**, 1010-1016.

James TW, Bohman R (1981) Proliferation of mitochondria during the cell cycle of the human cell line (HL-60). *J. Cell Biol.* **89**, 256-260.

Laird AK (1969) Dynamics of growth in tumours and normal organisms. *Natl. Cancer Inst. Monogr.* **30**, 15-28.

Lefkovits I, Waldmann H (1979) *Limiting dilution analysis of cells in the immune system*, Cambridge: Cambridge Univerisity Press.

McAdams HH, Arkin A (1999) It's a noisy business! Genetic regulation at the nanomolar scale. *Trends Genet.* **15**, 65-69.

Norton L (1985) Implications of kinetic heterogeneity in clinical oncology. *Semin. Oncol.* **12**, 231-249.

Posakony JW, England JM, Attardi G (1977) Mitochondrial growth and division during the cell cycle in HeLa cells. *J. Cell Biol.* **74**, 468-491.

Robin ED, Wong R (1988) Mitochondrial DNA molecules and virtual number of mitochondria per cell in mammalian cells. *J. Cell Physiol.* **136**, 507-513.

Rubin H (2005) Degrees and kinds of selection in spontaneous neoplastic transformation: an operational analysis. *Proc. Natl. Acad. Sci. USA* **102**, 9276-9281.

Vineis P (2003) Cancer as an evolutionary process at the cell level: an epidemiological perspective. *Carcinogenesis* **24**, 1-6.

Weinberg RA (1989) Oncogenese, antioncogenes, and the molecular basis of multistep carcinogenesis. *Cancer Res.* **49**, 3713-3721.

Wilson GD (2001) A new look at proliferation. *Acta Oncol.* **40**, 989-994.

FIGURE CAPTIONS

Fig.1 - Measuring the growth of tumour cells by microplate spectrophotometry. **A.** Mean absorption spectra measured at the indicated times for 60 wells containing an initial number of about 1000 cells (solid lines) and for 60 wells containing growth medium only (dashed lines). The SD is not shown for the sake of clarity, however the observed maximum CV was <2%. **B.** The growth of 6 cell populations from an initial cell density of 10000 cells/ml was measured by microplate spectrophotometry at the 730 nm wavelength and data averaged (black circles). For comparison, cell growth was measured in parallel for 6 independent cell cultures grown under the same experimental conditions by counting the cells at the microscope using the trypan-blue dye exclusion assay (open squares). **C.** The same spectrophotometric data shown in B are here compared to ATP measurements by the luciferine/luciferase method carried out on 6 independent cell cultures grown under the same experimental conditions. **D.** Four Molt3 cell clones were isolated in different wells of a microplate and their growth monitored *in situ* by spectrophotometry. The lines represent the best fits obtained with equation (1). The figure provides an example of the observed growth variability of clonal populations.

Fig.2 - Epigenetic basis of the growth variability of tumour cell clones. Molt3 cells were cloned and the growth rate of each clonal population was estimated by fitting experimental data with equation (1) as shown also in Fig.1D. The distribution of the growth rates measured for N=201 clonal populations is shown along with the mean±SD growth rate *r*. Six clones growing with the indicated rates ($r_p$), where the index *p*

denotes that the rate is the mean rate of the clonal population, were then selected and cells expanded in T25 culture flasks for 4 days. Two out of six clones showed a growth rate below the mean value observed for cloned Molt3 cells and the other 4 a value above. Then, cells from these populations were further cloned and the mean ± SD value measured for the indicated number of clonal populations obtained is reported. For statistical purposes, the growth rates measured for the clonal populations obtained from the 6 initially selected clones were grouped as indicated and the two distributions obtained are shown in the figure. These distribution are similar to that of the parent Molt3 populations. The mean growth rates do not significantly differ from that of the parent population (Student $t$ test, $P>>0.05$). Thus, "the growth rate" is not a genetically inheritable trait.

Fig.3 - Flow cytometry experiments for the analysis of mitochondria in Molt3 cells. **A.** Bi-parametric analysis carried out with mitochondria and nuclear DNA specific fluorescent probes. On average, the number of mitochondria in G2/M phase approximately doubles the value observed for cells in G1. **B.** Mitochondria isolated from Molt3 cells (upper panel) and mitochondria in intact cells (bottom panel) were loaded with a specific fluorescent probe and analysed. The fluorescence intensity was collected for both samples at the same sensitivity of the photomultiplier. The black-filled distributions of fluorescence intensity values show the data obtained for unlabelled mitochondria and cells.

Fig.4 - Distributions of the growth rates of Molt3 clones and fits with model equation (7). The distributions of the growth rates measured for Molt3 clones from 3 independent

experiments are shown (A, B and C). The values measured in these experiments were also grouped and analysed (right panel). In all figures, the line represents the best fit obtained with equation (7), the symbol N refers to the number of analysed clones, $k$ is the linear growth rate of mitochondria estimated by data fitting and $\sigma$ is the estimated standard deviation of the fluctuations of the maximum number of mitochondria around the mean number (see the text for details).

Fig.5 - Distributions of the growth rates of Molt3 $\rho^0$ cell clones. **A.** PCR analysis of $\rho^0$ cells. In each lane, 25 µl of the PCR products were loaded. Lanes are as follows: 1. nuclear β-actin gene fragment, normal cells; 2. mitochondria DNA-specific fragment, normal cells; 3. nuclear β-actin gene fragment, $\rho^0$ cells; 4. mitochondria DNA-specific fragment, $\rho^0$ cells. Traces of mitochondrial DNA are commonly visible in $\rho^0$ cells (see also Armand *et al.* 2004). Thus, to demonstrate that these cells are true $\rho^0$ functional assays must also be performed and in this paper we have monitored their dependence on uridine and pyruvate for survival. **B.** Distribution of the growth rates of $\rho^0$ cell clones. Data are plotted on the same x-axis scale as that used for normal cells for comparison (see Fig.2).

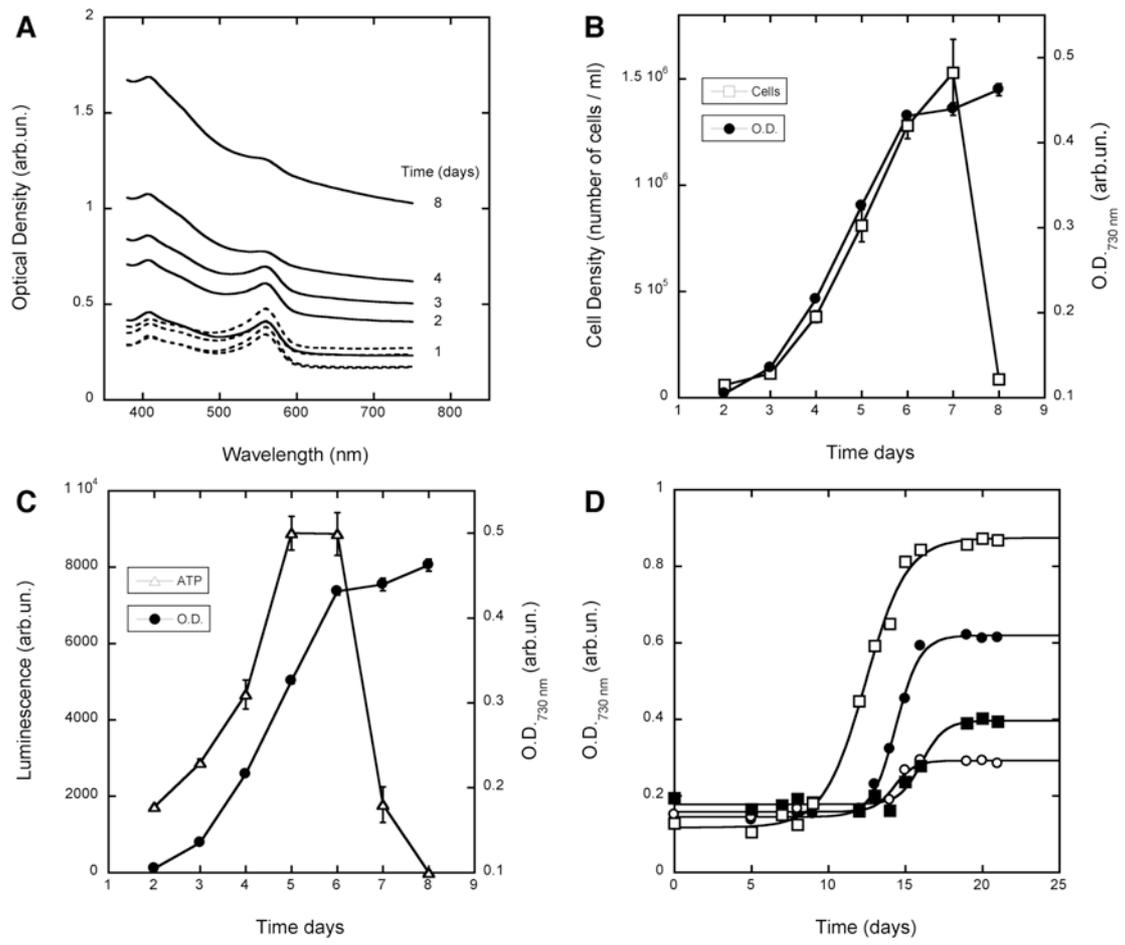

Fig.1

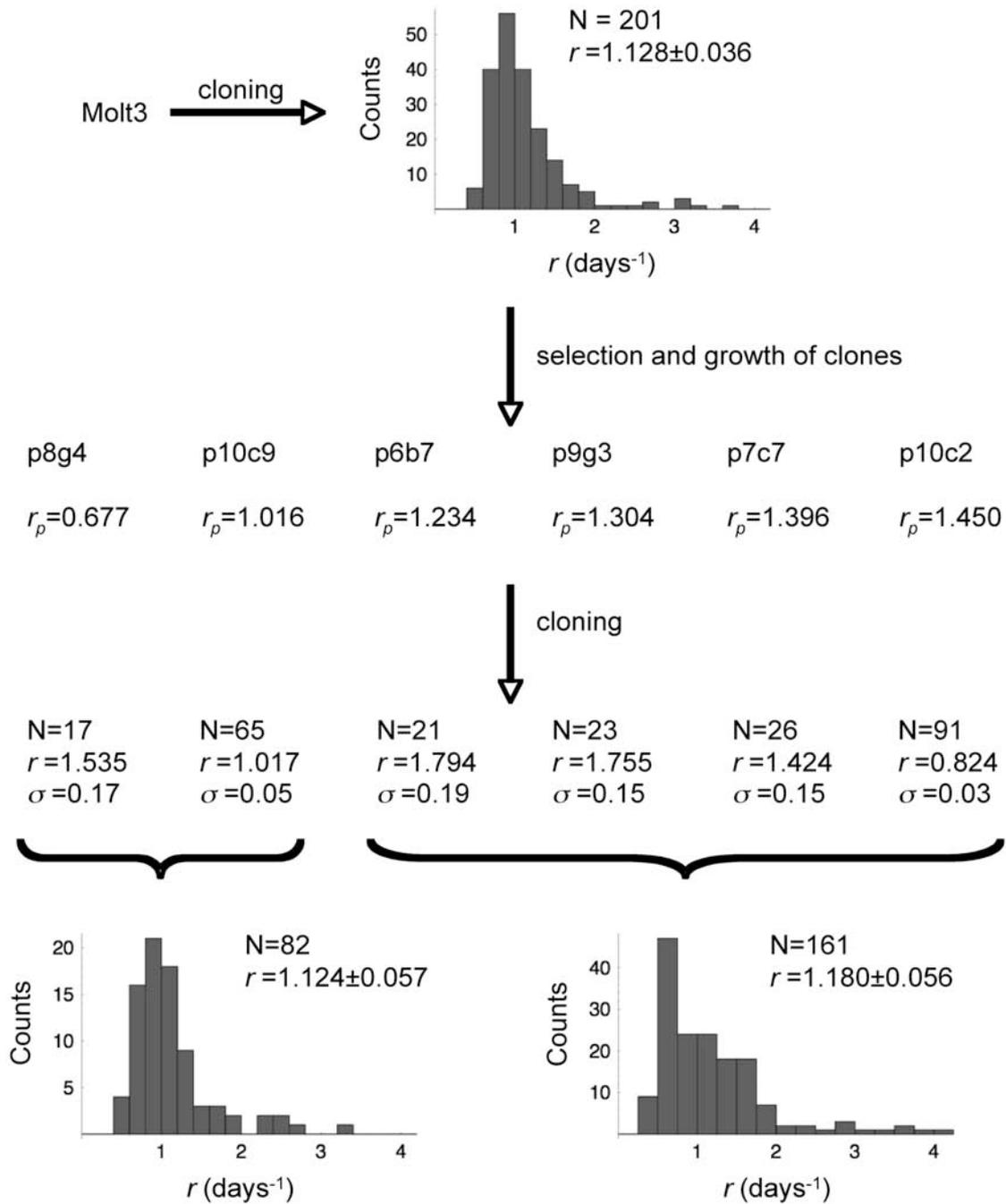

Fig.2

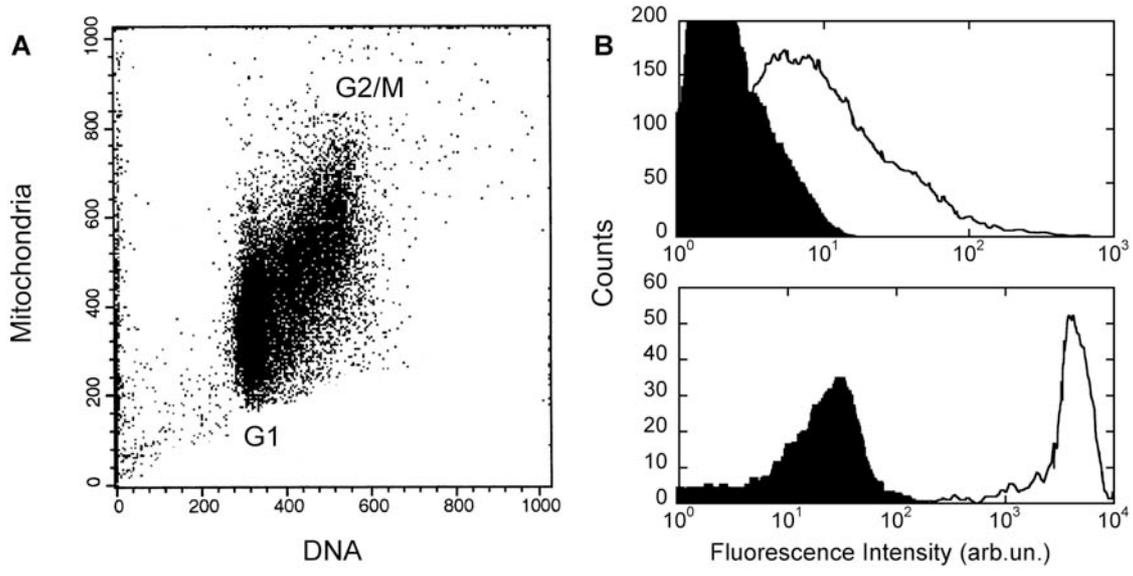

Fig.3

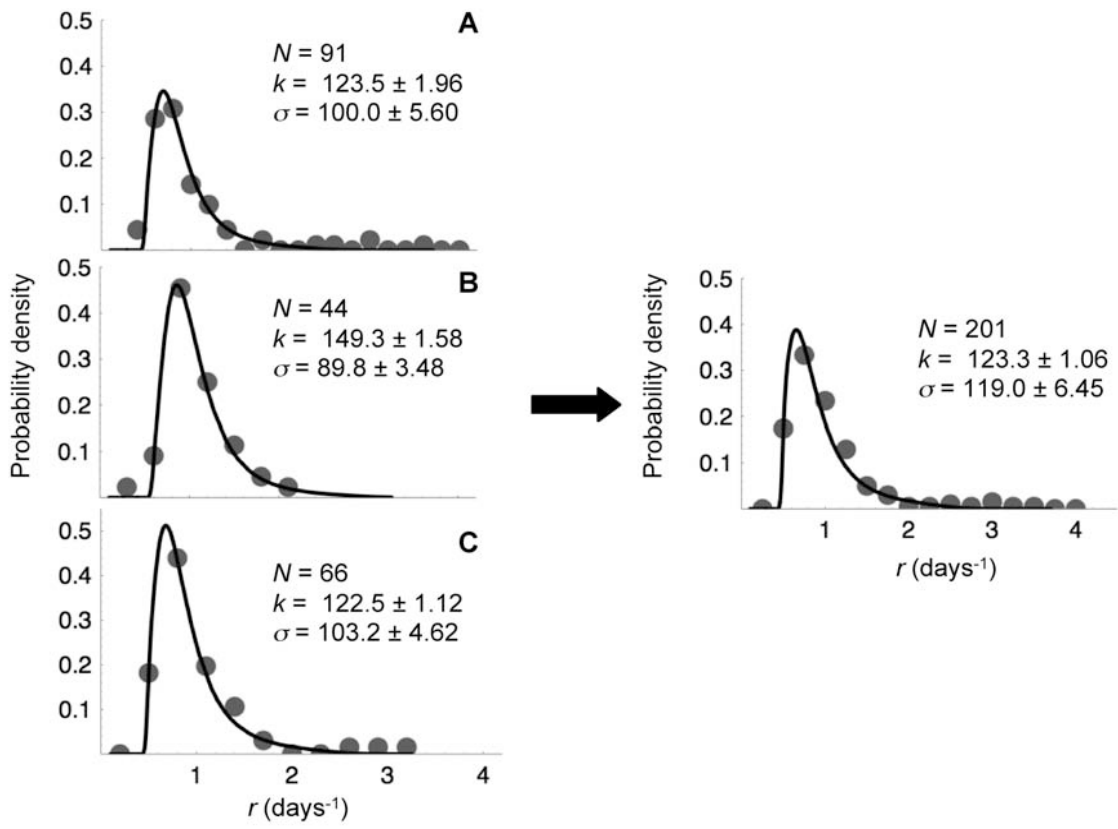

Fig.4

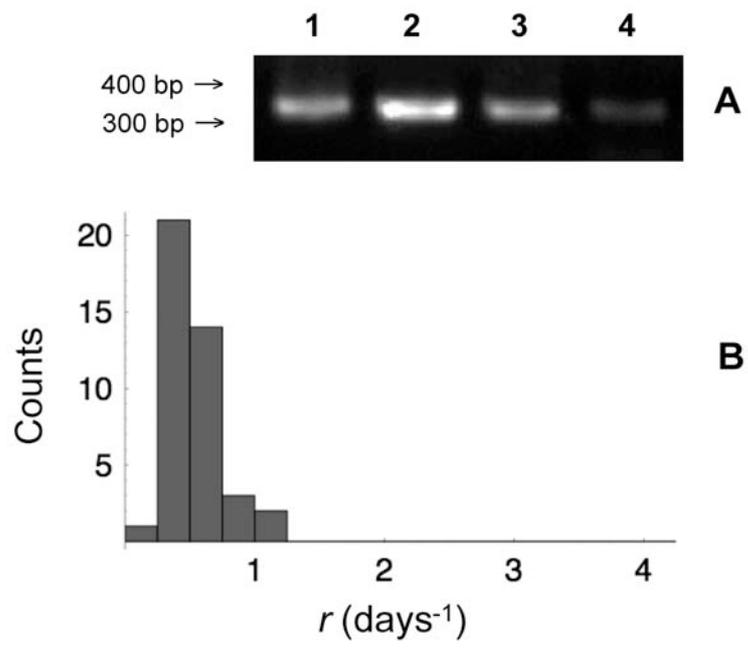

Fig.5